\documentclass[aps,prb,reprint,showpacs,superscriptaddress,longbibliography]{revtex4-2}
\usepackage{mathtools,amssymb,graphicx,units}

\usepackage[plainpages=false,pdfpagelabels,colorlinks=true,linkcolor=red,urlcolor=blue,citecolor=blue,pdftitle={Title},pdfauthor={},pdfdisplaydoctitle=true,pdfduplex=DuplexFlipLongEdge]{hyperref}
\usepackage{verbatim}
\usepackage[english]{babel}
\usepackage{color}

\usepackage[normalem]{ulem}

\begin{document}

\title{Interplay of interactions, disorder, and topology in the Haldane-Hubbard model}
\author{Tian-Cheng Yi}
\affiliation{Beijing Computational Science Research Center, Beijing 100193, China}
\author{Shijie Hu}
\affiliation{Beijing Computational Science Research Center, Beijing 100193, China}
\author{Eduardo V. Castro}
\affiliation{Centro de F\'\i sica das Universidades do Minho e Porto, Departamento de F\'\i sica e Astronomia, Faculdade de Ci\^encias, Universidade do Porto, 4169-007 Porto, Portugal}
\affiliation{Beijing Computational Science Research Center, Beijing 100193, China}
\author{Rubem Mondaini}
\email{rmondaini@csrc.ac.cn}
\affiliation{Beijing Computational Science Research Center, Beijing 100193, China}

\begin{abstract}
We investigate the ground-state phase diagram of the \textit{spinless} Haldane-Hubbard model in the presence of quenched disorder, contrasting results obtained from both exact diagonalization and the density matrix renormalization group, applied to a honeycomb cylinder. The interplay of disorder, interactions, and topology gives rise to a rich phase diagram and, in particular, highlights the possibility of a disorder-driven trivial-to-topological transition in the presence of finite interactions. That is, the topological Anderson insulator, demonstrated in noninteracting settings, is shown to be stable in the presence of sufficiently small interactions before a charge density wave Mott insulator sets in. We further promote a finite-size analysis of the transition to the ordered state in the presence of disorder, finding a mixed character of first- and second-order transitions in finite lattices, tied to the specific conditions of disorder realizations and boundary conditions used.
\end{abstract}
\maketitle

\section{Introduction}\label{sec:intro}
Topological phase transitions have been intensively studied over the last few decades, especially after the discovery~\cite{Klitzing1980} and theoretical understanding~\cite{Thouless1982,Niu1985} of the quantum Hall effect. Among the many interesting properties of topological systems what stands out is their remarkable resilience to the introduction of disorder, provided it does not break any fundamental symmetry of the Hamiltonian~\cite{Xiao2010,Qi2011}. On this front, even more peculiar phenomena have been demonstrated, such as the case in which an otherwise topologically trivial system could turn nontrivial in the presence of sufficiently large disorder. These systems are dubbed topological Anderson insulators (TAIs)~\cite{Li2009,Jiang2009} and were shown to be a consequence of the renormalization of the trivial mass of the models by the presence of (small) disorder, rendering a trivial-to-topological phase transition possible~\cite{Groth2009}. Subsequent studies demonstrated similar phenomenology in the Haldane model~\cite{Xing2011,Song2012,Goncalves2018,Sriluckshmy2018}, in the Kane-Mele model~\cite{Orth2016}, and in models for quantum wells~\cite{Yamakage2011,Prodan2011,Zhang2012,Chen2012} in three-dimensional topological systems~\cite{Guo2010}, culminating in its experimental observation in ultracold atoms~\cite{Meier2018} and photonic waveguides~\cite{Stutzer2018}.

In turn, the study of \textit{interacting} topological systems is also diverse~\cite{Hohenadler2013} and, in some cases, controversial. While antiferromagnetic topological insulating states~\cite{Mong2010, Fang2013, Yoshida2013, Miyakoshi2013} and even interaction-driven topological Mott insulators in otherwise topologically trivial models~\cite{Raghu2008, Wen2010, Weeks2010, Budich2012, Dauphin2012, LeiWang2012, Ruegg2011, Yang2011} have been argued to exist, a consensus cannot be reached when using unbiased numerical calculations~\cite{Garcia_Martinez2013, Daghofer2014, Motruk2015, Capponi2015, Scherer2015}. The concomitant appearance of nontrivial topology and the formation of a local charge or magnetic order parameters has also been shown to be nonexistent in a variety of fermionic models~\cite{Varney2010, Varney2011, Yamaji2011, Zheng2011, Yu2011, Hohenadler2011, Griset2012, Hohenadler2012, Reuther2012, Budich2013, Laubach2014, Amaricci2015, Wu2016, Shao2021}.

Building on these results, we aim to study the much less explored combined effects of disorder and interactions in topological systems, which render a systematically richer physics. Along these lines, earlier studies investigated such interplay using approximative methods, such as Hartree-Fock~\cite{Sinova2000}, perturbative renormalization group methods~\cite{Ostrovsky2010, Wang2017}, and random phase approximation~\cite{Makogon2010}. More recent studies followed our approach of making use of unbiased numerical methods but mostly focused on quasi-one dimensional systems~\cite{Zhang2021, Li2021}. 

Here, by using a combination of unbiased approaches, exact diagonalization (ED) in small clusters, and the density matrix renormalization group (DMRG) in cylinders, we unveil the phase diagram of the disordered Haldane-Hubbard model. Among our results, we show that the TAI phase is also manifest in the presence of interactions in regimes in which they are not sufficiently strong to trigger a topologically trivial Mott insulating phase. In addition, we perform a careful finite-size analysis of the charge density wave (CDW) Mott insulating transition, finding that at small disorder amplitudes, first- and second-order phase transitions may occur in finite systems with specific boundary conditions, separating the topologically trivial and nontrivial phases.

\begin{figure*}[ht!]
\centering
\includegraphics[width=1\textwidth]{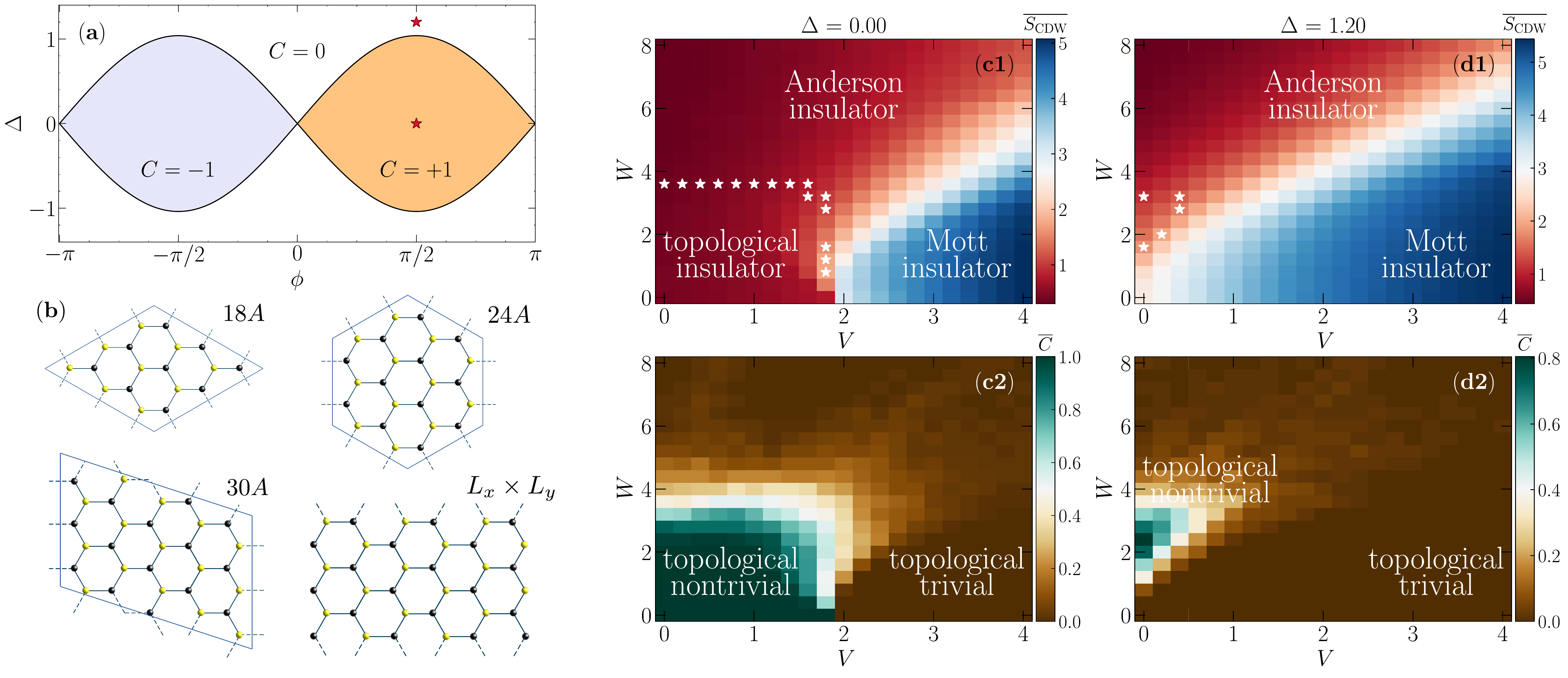}
\caption{(a) Phase diagram of the noninteracting Haldane model, with stars depicting the set of parameters in which the analysis in the presence of both disorder and interactions was performed. (b) The lattice clusters used in the ED and DMRG calculations, with dashed lines emphasizing the periodic boundary conditions used; in the $L_x \times L_y$ cylinders used in the DMRG (here a $6\times6$ is sketched), open boundary conditions are employed along its longitudinal direction. (c1) and (c2) Phase diagrams depicting the disorder-averaged charge density wave structure factor $\overline{S_{\rm CDW}}$ and Chern number, respectively. (d1) and (d2) The same as in (c1) and (c2), but for the case of a finite trivial mass, $\Delta=1.2$. Stars in (c1) an (d1) delimit the topological region with a working definition, where $\overline{C} \simeq0.5$. The number of disorder realizations is 100, and the results are for the $24A$ lattice.
}
\label{fig:fig_1}
\end{figure*}

\section{Model and quantities} \label{sec:model}
We consider a Hamiltonian $\hat {\cal H}$ that is a combination of the Haldane model on a honeycomb lattice~\cite{Haldane1988},
\begin{eqnarray}
&&{\cal \hat H}_{\rm Haldane}
\nonumber\\
&=&-t_1\sum_{{\langle i,j\rangle}} \left( \hat c^\dag_i \hat c^{\phantom\dag}_j + {\rm H.c.} \right)
%\nonumber\\&&
-t_2\sum_{{\langle\langle i,j\rangle\rangle}} \left(e^{{\rm i} \phi_{ij}} \hat c^\dag_i \hat c^{\phantom \dag}_j + {\rm H.c.} \right)
\nonumber\\&&+\Delta \sum_i (-1)^{i} \hat n_i^{\phantom\dag}
%\nonumber\\&&
,
  \label{eqn:haldaneham}
\end{eqnarray}
and both contact interactions and onsite disorder,
\begin{equation}
\hat {\cal H}_V + \hat {\cal H}_W  = V \sum_{{\langle ij\rangle}} \hat n_i \hat n_j + \sum_i W_i \hat c^\dag_i \hat c^{\phantom\dag}_i.
\end{equation}
Here, $\hat c_i^\dag$ ($\hat c_i^{\phantom\dag}$) represents the fermion creation (annihilation) operator at site $i$, and $\hat n_i =\hat c^\dag_i \hat c_i^{\phantom\dag}$ is the corresponding number operator. $t_1$ ($t_2$) is the nearest-neighbor (next-nearest-neighbor) hopping amplitude, and $\Delta$ is the staggered potential responsible for breaking the symmetry between the two sublattices of a honeycomb lattice. The next-nearest-neighbor hopping term has a complex phase $\phi_{ij} = +\phi (-\phi)$ for counter-clockwise (clockwise) hoppings; $V$ describes the magnitude of a repulsive nearest-neighbor interaction, and $W_i$ is disordered on-site energy. In this case, we choose $W_i$ from a uniform random distribution [$-W$,$W$], emulating a fully disordered system. In what follows, we set $t_2=0.2t_1$ and $t_1=1$ as the energy scale.

By employing ED in finite lattices from $N_s = 18$ to 30 sites and DMRG in cylinders comprising up to 72 orbitals, we are able to characterize the low-lying spectral properties that govern the topological behavior of $\hat{\cal H} = {\cal \hat H}_{\rm Haldane} +\hat {\cal H}_V + \hat {\cal H}_W $, focusing the investigation on half filling, i.e., $N_e\equiv \sum_i \langle \hat n_i\rangle/N_s = N_s/2$. For both types of simulations, we carefully select the finite lattice structures such that they contain the $K$ high-symmetry point as a valid momentum value. This is fundamental for assessing the low-energy properties of the Hamiltonian, in particular the first order phase transition separating topological and trivial regions with vanishing disorder amplitude~\cite{Varney2010,Varney2011,Shao2021,Shao2021b}. A representation of the clusters used within both approaches is given in Fig.~\ref{fig:fig_1}(b). In the DMRG calculations, we use cylinders with a fixed circumference $L_y=6$ and various lengths $L_x=4$ to $12$ ($N_s=L_x L_y$); in addition, the truncation dimensions are set at 1024, and five sweeps are performed for an arbitrary set of parameters in order to guarantee the accuracy. In ED, we employ Krylov-Schur methods~\cite{Petsc,Slepc} to extract the ground state and a few excited states of $\hat {\cal H}$.

Our starting point is the celebrated phase diagram of $\hat {\cal H}_{\rm Haldane}$ in the $\Delta$ vs $\phi$ plane [Fig.~\ref{fig:fig_1}(a)]. The lines at $\Delta = \pm 3\sqrt{3}t_2\sin\phi$, the gap closing condition~\cite{Haldane1988}, separate the three possible regions with different Chern numbers. In our study, we focus on the two points marked in Fig.~\ref{fig:fig_1}(a), $\phi = \pi/2$ with $\Delta = 0$ or $1.2$, and subsequently study the effects of disorder and interactions. That is, we investigate how the original topological Chern insulator state (at $\Delta=0$) is affected by these two "knobs". Furthermore, we also study the case in which a topological Anderson insulator for $\Delta\neq0$~\cite{Goncalves2018} is resilient to the contact interactions.

To quantify the topological nature of the ground state, in ED, we compute the Chern number $C$ using a discretized form of the integration of the Berry curvature~\cite{Fukui05,Varney2010,Varney2011,Shao2021},
\begin{equation}
    C = \int \frac{d\phi_x d\phi_y}{2 \pi { i}} \left( \langle\partial_{\phi_x}
      \Psi^\ast | \partial_{\phi_y} \Psi\rangle - \langle{\partial_{\phi_y}
      \Psi^\ast | \partial_{\phi_x} \Psi\rangle} \right),
      \label{eq:Chern}
\end{equation}
after twisted boundary conditions $\{\phi_x,\phi_y\}$ are employed. Details of this discretization are explained in Appendix~\ref{appendixA}. In the noninteracting limit, we employ a coupling matrix approach~\cite{Zhang2013}, which allows accurate computation of the topological invariant using a single boundary condition and is particularly useful within disorder settings~\cite{Goncalves2018}.

At large repulsive interaction strengths, the topological insulator gives way to a trivial CDW insulator through a topological
phase transition~\cite{Varney2010,Varney2011}. In the limit $V \to\infty$, the ground state will be a perfect CDW, in which one of the two sublattices is occupied while the other is empty, leaving lattice translational symmetry intact but breaking reflection symmetry~\cite{PhysRevB.75.174301}. To characterize it, we compute the ${\bf k}=0$ CDW structure factor~\cite{Varney2010,Varney2011,Shao2021}
\begin{align}
  \label{eqn:struct}
  S_{\rm CDW}  &\equiv \frac{1}{N}\sum_{i,j} C({\bf r}_i - {\bf r}_j)
\end{align}
with density correlations
\begin{align}
  C({\bf r}_i - {\bf r}_j) &= \langle (\hat n_i^a - \hat n_i^b) (\hat n_j^a - \hat n_j^b) \rangle,
\end{align}
where $\hat n_i^a$ and $\hat n_i^b$ are the number operators on sublattices $a$ and $b$
in the $i$-th unit cell, respectively. Here, $N$ is the total number of unit cells ($N=N_s/2$).

We also compute the fidelity susceptibility~\cite{Zanardi06,CamposVenuti07,Zanardi07,You2007}, 
\begin{equation}
    \chi = \frac{2}{N} \frac{1 - \langle \psi_0(V)|\psi_0(V+dV)\rangle}{dV^2},
\end{equation}
which is a natural way to pinpoint when a quantum phase transition takes place in the regime of parameters of the interest~\cite{Yang07,Varney2011,Jia11,Mondaini15} by making no direct assumptions about the order parameter associated with it. Here the overlap of the ground-state wave functions with a small interaction difference $dV=10^{-3}$ is used to understand the effects of interactions in suppressing the topological character of $|\psi_0\rangle$ and its competition with a disordered Anderson phase. In fact, due to the fluctuating nature of this quantity in the presence of disorder, we will compute its typical value $\chi_{\rm typ.} \equiv \exp{\left(\overline{ \ln \chi}\right)}$, where $\overline{(\cdot)}$, here and elsewhere, denotes the disorder averaging.

Finally, as direct verification of when the Anderson insulating phase takes place, we compute the single-particle density of states (DOS), resolved in its electron and hole channels,
\begin{align}
{\cal N}(\omega) &= {\cal N}^+(\omega) +  {\cal N}^-(\omega) \nonumber \\ 
                 =\frac{1}{N_s}\sum_i\Big\{&\sum_{n} |\langle \psi_0| \hat c^\dagger_i| \psi_n^{N_e-1}\rangle|^2 \delta\left[\omega + (E_n^{N_e-1}-E_0)\right] \nonumber \\ 
                 +&\sum_n |\langle \psi_0| \hat c_{i}| \psi_n^{N_e+1}\rangle|^2 \delta\left[\omega - (E_n^{N_e+1}-E_0)\right]\Big\},
\end{align}
where $\{E_n^{N_e\pm1},\psi_n^{N_e\pm1}\}$ is an eigenpair of the Hamiltonian with an added or removed electron. When we compute it, we truncate the sum to use up to 100 eigenpairs, further employing standard disorder averaging. 

\section{Results}
We start by displaying our main results in Figs.~\ref{fig:fig_1}(c) and \ref{fig:fig_1}(d), obtained for the highly symmetric $24A$ cluster and qualitatively verified on the remaining ones. In Figs. 1(c1) and 1(c2), in the absence of the trivial mass term $\Delta$, we directly notice that the topological insulating phase is disjoint from the Mott insulating one; that is, in the presence of disorder, the Chern insulating  with $C = 1$ does not overlap with the one associated with a finite local order parameter, signified by a large $S_{\rm CDW}$. This generalizes a known result for this system at vanishing disorder~\cite{Varney2010,Varney2011}. At large disorder amplitudes, both the (disorder averaged) Chern number and CDW structure factor tend to zero, which leads to the identification of an Anderson insulating phase. For $V=0$, the transition perfectly agrees with known results previously obtained for much larger systems ($N_s \simeq 5\times 10^3$)~\cite{Goncalves2018}.

For a finite staggered potential, on the other hand, a trivial CDW phase emerges (that is, not necessarily associated with the contact interactions). At $\Delta=1.2$ the system is topologically trivial [see Fig.~\ref{fig:fig_1}(a)] and remains so with increasing $V$. In contrast, when we include disorder, we observe the formation of a $W$-driven topologically nontrivial state, the topological Anderson insulator in the Haldane model~\cite{Xing2011,Song2012,Goncalves2018,Sriluckshmy2018}. What our results further advance is that this phase is robust to the presence of interactions, provided they are not sufficiently large to give rise to the trivial CDW Mott insulator [see Fig.~\ref{fig:fig_1}(d1) and 1(d2) and the Appendix~\ref{appendixC} for a smooth variation of the staggered potential].

In what follows, we address four specific points: (i) the details of the TAI phase, (ii) the finite-size analysis of the CDW insulating transition and the drop in the Chern number, (iii) characterization via the fidelity metric, and (iv)  inference of the trivial Anderson insulating phase via the single-particle density of states.

\subsection{Details of the interacting topological Anderson insulator}
\begin{figure}[t]
\centering
\includegraphics[width=0.90\columnwidth]{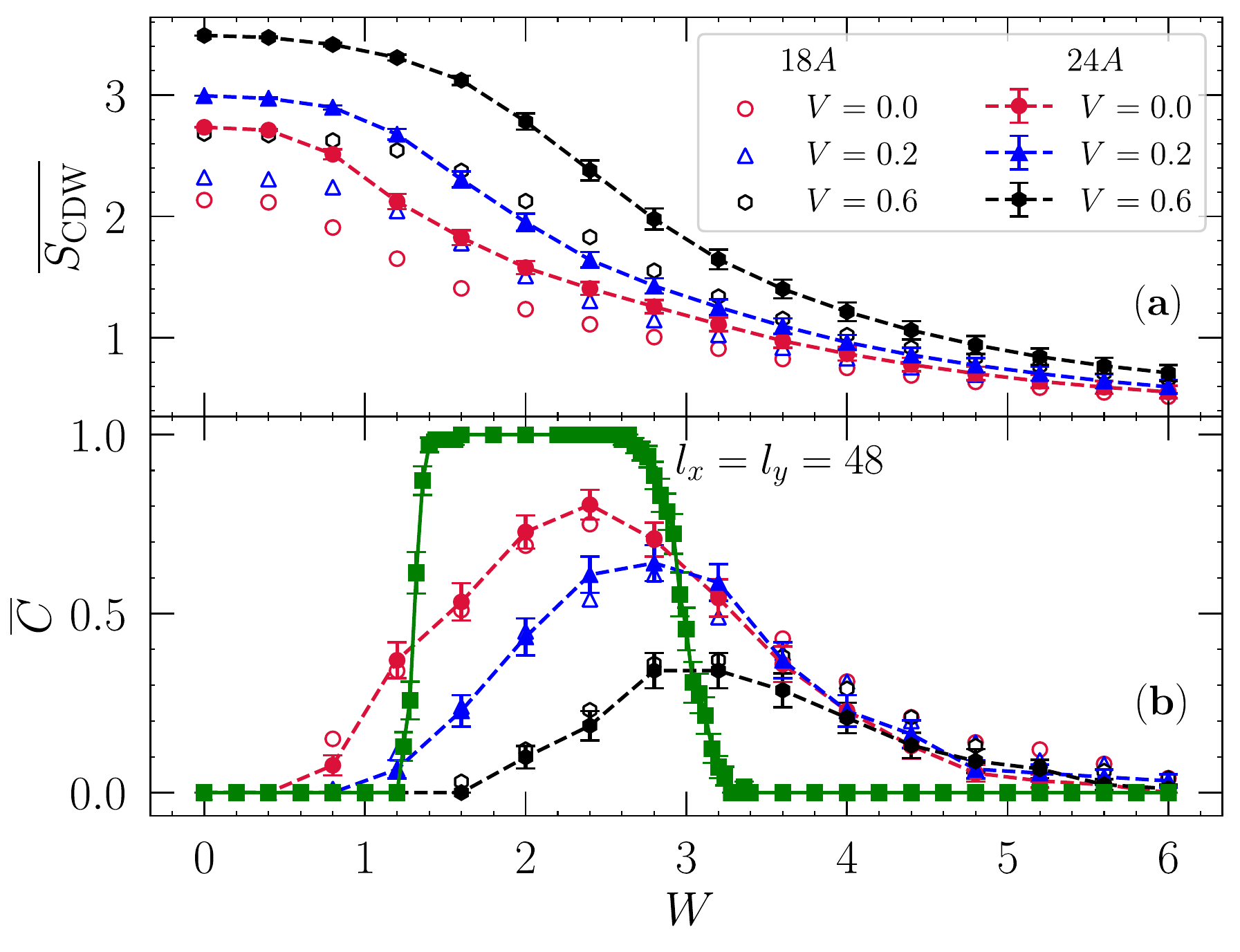}
\caption{(a)The disorder-averaged structure factor $\overline{S_{\rm CDW}}$ and (b) Chern number $\overline C$ for increasing $W$ for different $V$ via ED for 100 realizations for $18A$ (open markers) and $24A$ (solid markers). Here we fix the staggered term at $\Delta=1.2$. The nonmonotonic behavior of $\overline C$ is indicative of the interacting topological Anderson insulator phase and can be seen to smoothly depart from the noninteracting results [square markers in (b) for a $48\times 48$ lattice], with small size effects that point to growth with system size at intermediate disorder amplitudes. Error bars for the $18A$ results are omitted for clarity.}
\label{fig:int_tai}
\end{figure}

The interacting TAI we just described can be seen in more detail in Fig~\ref{fig:int_tai}, where the disorder-averaged Chern number and the corresponding CDW structure factor are displayed at a few finite values of $V$ while sweeping the disorder amplitude. Here we compare two clusters, $18A$ and $24A$, showing that with increasing $W$, $\overline C$ exhibits a nonmonotonic behavior, increasing with intermediate values of disorder but decreasing after reaching the trivial Anderson insulating phase. We notice that such behavior occurs for values of interactions which are sufficiently small; otherwise, they give rise to the trivial CDW Mott insulator, which in our model, precludes the manifestation of a topological phase. Nonetheless, the average Chern number can be seen to smoothly depart from its noninteracting limit as the interacting strength is increased. The finite-size effects, which are small, and whose fate is difficult to definitively determine on approaching the thermodynamic limit, are suggestive of its quantization in that limit. The formation of the interacting TAI phase with growing $\Delta$ can be seen in  Appendix~\ref{appendixC}.

\subsection{Finite-size effects and Chern to Mott insulator transition}

\begin{figure}[t]
\centering
\includegraphics[width=0.90\columnwidth]{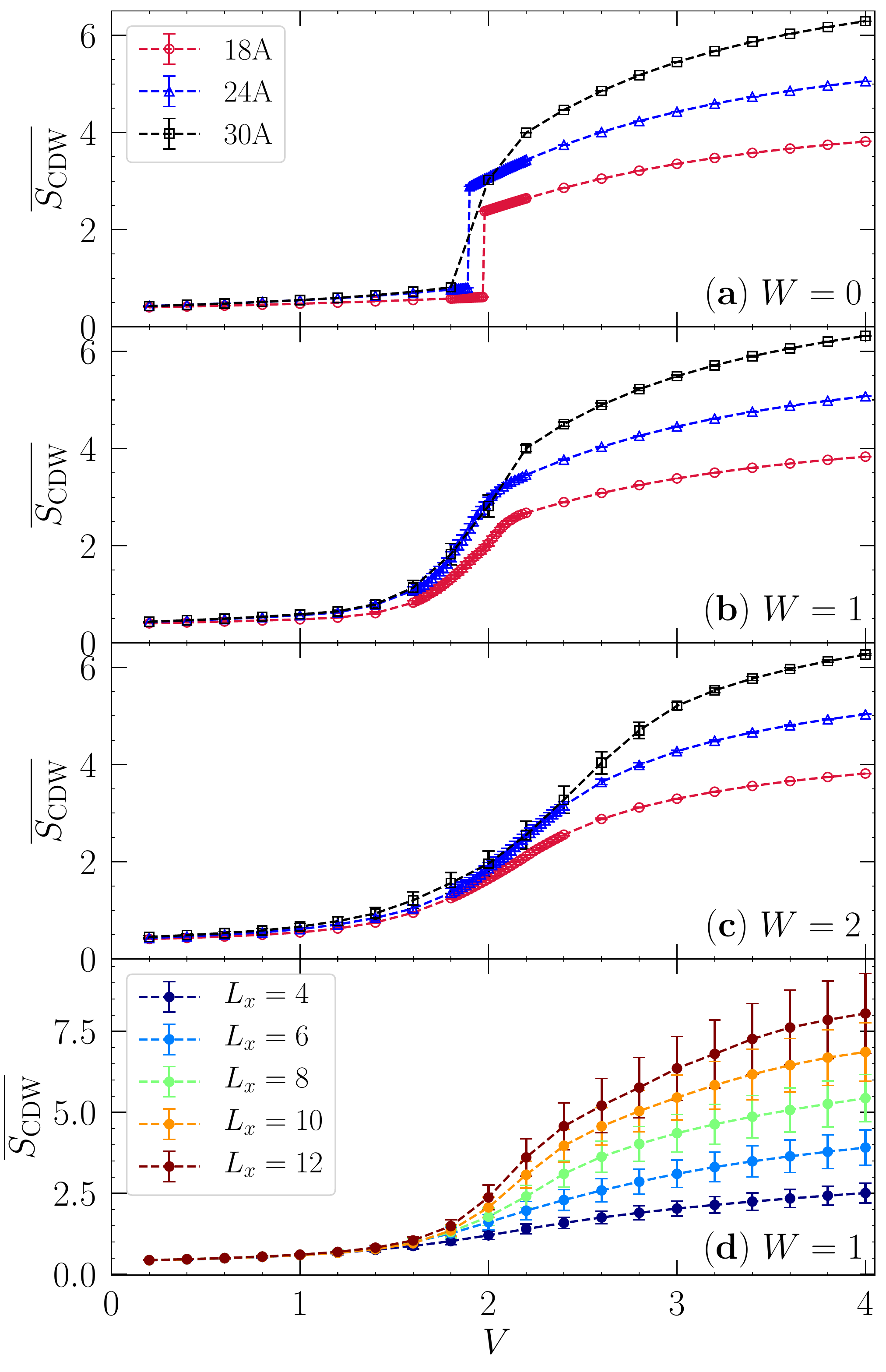}%data030201_merge0618_2_W0_N18_mergeW030_W0upW1W2
\caption{(a)-(c)The structure factor $\overline{S_{\rm CDW}}$ with respect to $V$ for various cluster sizes and different disorder strengths $W$ using ED with 100 realizations. (d) The same as in (a)-(c) for cylinders with growing dimensions extracted from DMRG calculations and fixed $W=1$. All data for both methods refer to $\Delta=0.0$.}
\label{fig:scdw}
\end{figure}

\begin{figure}[t]
\centering
\includegraphics[width=0.95\columnwidth]{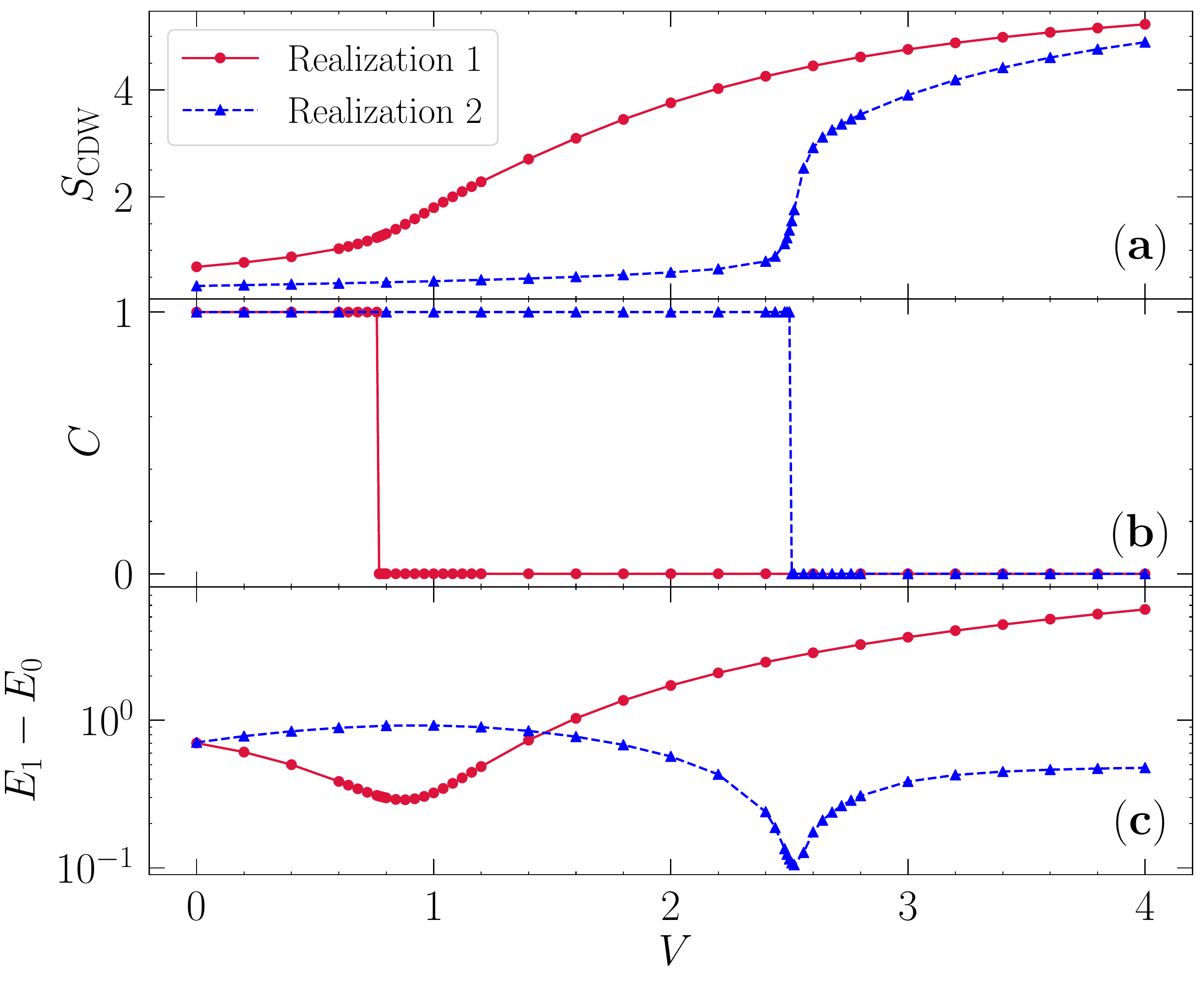}
\caption{(a) CDW structure factor, (b) Chern number, and (c) the many-body gap at periodic boundary conditions, $\phi_x=\phi_y=0$, for two different realizations for $\Delta=0$, $W=2$ and growing interactions. Here the cluster used is $24A$.
}
\label{fig:one_realz}
\end{figure}

\begin{figure}[t]
\centering
\includegraphics[width=0.90\columnwidth]{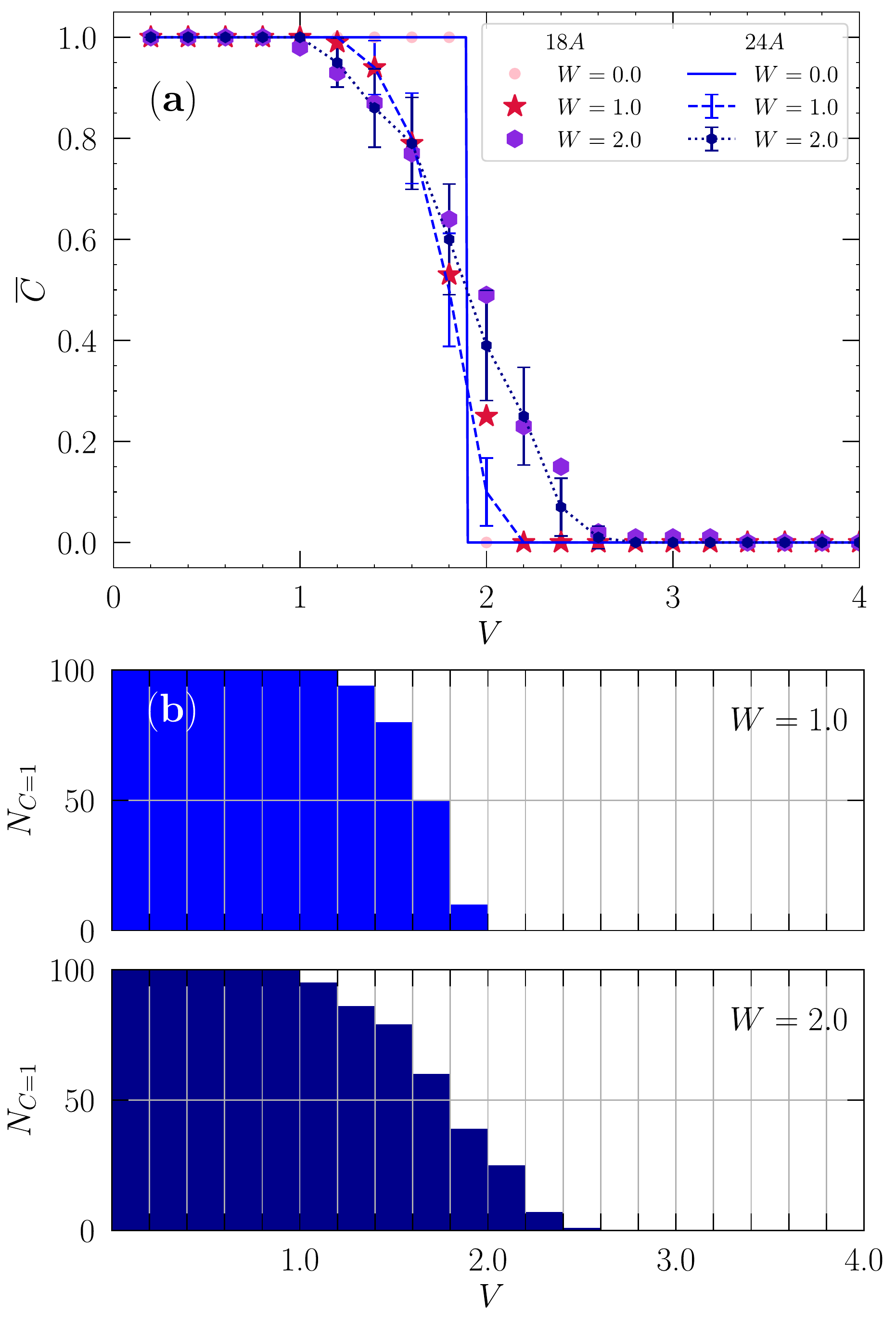}
\caption{The Chern number dependence on the interactions $V$ for different $W$ via ED. (a) The disorder-averaged value $\overline C$, comparing results for lattices $18A$ and $24A$; the clean result ($W=0$) in $24A$ is represented by a solid line. Error bars are displayed only at the largest cluster for clarity. (b) Histograms of the Chern number at two different disorder amplitudes as $V$ is increased on the $24A$ cluster. All data are obtained with $\Delta = 0$, using 100 disorder realizations.
}
\label{fig:C_FSE}
\end{figure}

Hereafter we focus on the $\Delta=0$ case. We start by noticing in Fig.~\ref{fig:scdw} that the known first-order phase transition giving rise to the Mott insulating phase in the clean case~\cite{Varney2010,Varney2011} is replaced by the typical second-order one when  sufficient disorder is included: the averaged structure factor smoothly interpolates between its $O(1)$ behavior at small interaction strengths and its extensive nature within the ordered phase. These results [Figs.~\ref{fig:scdw}(a)-\ref{fig:scdw}(c)], which were obtained via ED, are also confirmed by DMRG calculations using larger lattice sizes [Fig.~\ref{fig:scdw}(d)]. Here we fix the cylinder's transverse dimension at $L_y=6$ while systematically changing its length. 

%Although suggestive, a possible scaling analysis to determine the universality class of the transition turned out inconclusive. 
The observed trend of the system-size dependence of ${S_{\rm CDW}}$ raises the possibility of a proper scaling analysis to determine the universality class of the transition, which turned out to be inconclusive. One reason is that at small disorder amplitudes, the curves for $S_{\rm CDW}$ at a given disorder instance $\{W_i\}$, while the interaction magnitudes are swept can yield either a typical first-order (jump) or second-order (continuous) transition, with the former more likely to occur at small disorder values. This contrast is explicitly displayed in Fig.~\ref{fig:one_realz}(a), 
where we fix $W=2$ while comparing the $S_{\rm CDW}$ results for individual disorder realizations. Although both curves display a continuous transition from small to large structure factors, in some realizations this evolution is rapid. If we look at the corresponding many-body gap $E_1-E_0$ [Fig.~\ref{fig:one_realz}(c)] \textit{with periodic boundary conditions}, we notice that although it displays a minimum at these locations, it never closes. On the other hand, the Chern number in Fig.~\ref{fig:one_realz}(b) changes at these values of interaction. 

The interpretation that explains these results is that although the gap does not close for $(\phi_x,\phi_y) = (0,0)$, resulting in a smooth evolution of $S_{\rm CDW}$, it does change at some twisted boundary conditions $(\phi_x,\phi_y)$, which are sensed by the Berry curvature, ultimately changing the Chern number. Thus, the closer the actual gap closing occurs to $(\phi_x,\phi_y) = (0,0)$, the faster the change in $S_{\rm CDW}$ will be. Nonetheless, as further elaborated in Appendix~\ref{appendixB}, all boundary conditions are equivalent in the thermodynamic limit. As a result, if the topological character of the many-body wave function changes, either via disorder or via interactions, the gap must close, thus resulting in a first-order phase transition in this regime.

\begin{figure}[t]
\centering
\includegraphics[width=0.9\columnwidth]{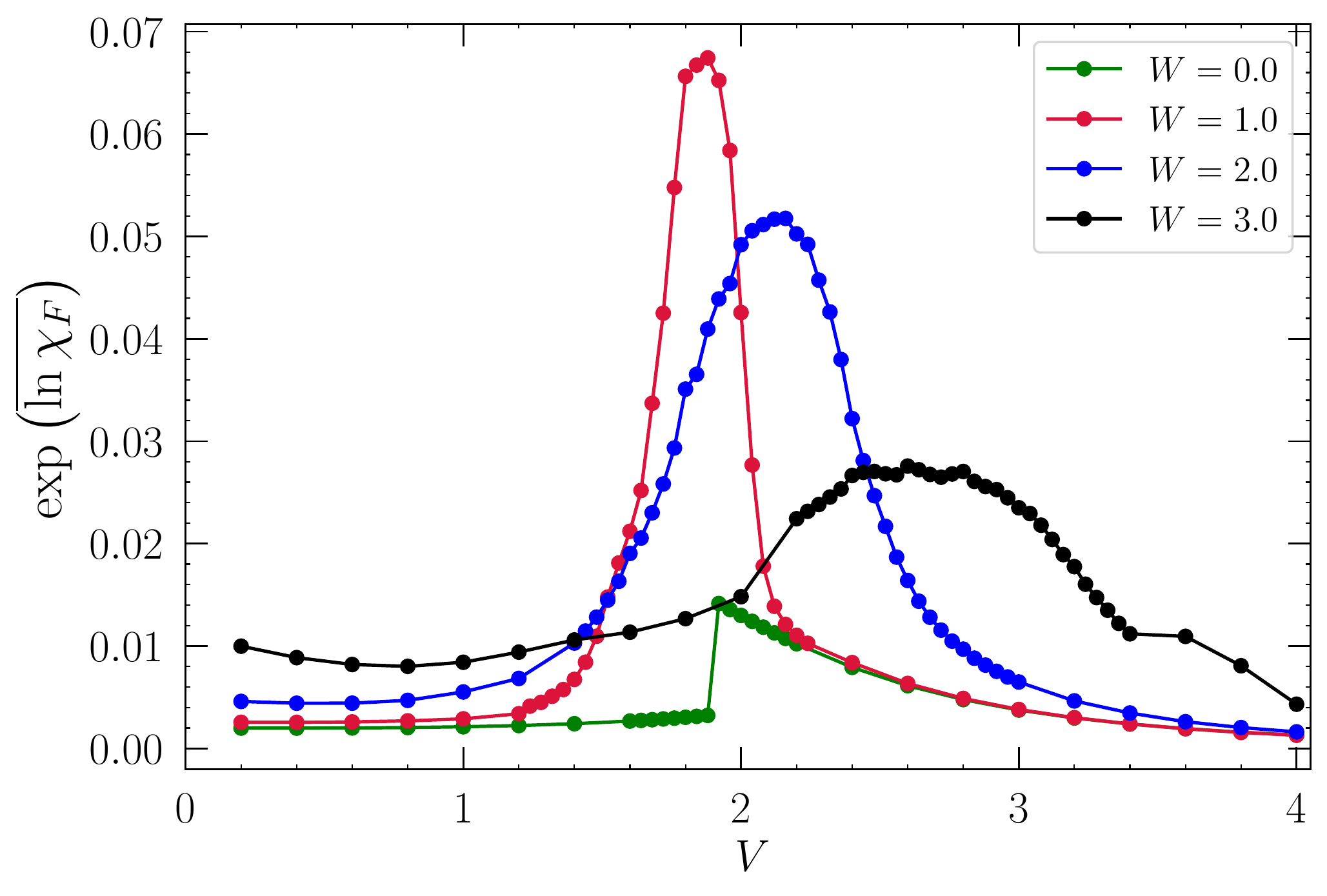}
\caption{Typical fidelity susceptibility $\exp(\overline{\ln\chi_F})$ with respect to $V$ for different $W$ via ED for $\Delta=0$ in lattice $24A$. The peak position closely locates the onset of the Mott-CDW phase in the phase diagram shown in Fig.~\ref{fig:fig_1}(c1).} 
\label{fig:fid_susc}
\end{figure}

In the phase diagrams in Figs.~\ref{fig:fig_1} (c1) and 1(c2), we notice that at small $W$, the Chern insulating phase directly neighbors the Mott one, and there are a large number of disorder realizations in which they follow the first-order phase transition separating them, as originally obtained in the clean case~\cite{Varney2011}. At large disorders, on the other hand, the topologically nontrivial phase is more unlikely to occur; therefore, a continuous phase transition separates the Anderson and Mott insulating phases. Given the symmetries of the Hamiltonian, this transition is likely to pertain to the chiral Ising universality class~\cite{Herbut2006,Herbut2009}.

In addition to the charge ordering, we further analyze the finite-size results of the Chern number in Fig.~\ref{fig:C_FSE}(a). While $C$ is a step function in the clean regime when the interaction strength is being swept, its disorder-averaged counterpart, $\overline{C}$, becomes a smooth interpolation between values of 0 and 1 at finite $W$. In the case of individual realizations, however, the Chern number is always quantized [see Fig.~\ref{fig:one_realz}(b)], and the continuous change is a result of the averaging procedure. The histograms in Fig.~\ref{fig:C_FSE}(b) emphasize the nature of this continuous change, and a possible definition of a topological transition could be given at the threshold where half of the disorder realizations result in a finite Chern number, as used in Figs.~\ref{fig:fig_1}(c1) and ~\ref{fig:fig_1}(d1) to draw the stars delimiting the topological region. Last, another important aspect is the overall small size effects in $\overline{C}$, which become evident when comparing lattices with $N_s=18$ and 24 in Fig.~\ref{fig:C_FSE}(a) and were similarly seen for the case of finite $\Delta$ [Fig.~\ref{fig:int_tai}(b)]. It is possible, however, that in approaching the thermodynamic limit, this continuous evolution becomes increasingly sharper, recovering quantization even after the disorder average.

\subsection{Fidelity susceptibility}

The ability of the fidelity susceptibility to infer phase transitions by directly checking how different ground states with marginally different parameters of the Hamiltonian are allows one to use it as a secondary tool to corroborate the phase diagrams in Fig.~\ref{fig:fig_1}. In the case of first-order phase transitions, this quantity diverges, $F\propto(dV)^{-2}$, while it displays an extensive (for the system size) peak when crossing a continuous one.

In the presence of a finite $W$, the fidelity metric can be a less accurate proxy due to noisy behavior after the disorder average, which, as we have seen, can mix first- and second-order phase transitions in a finite lattice. For example, by fixing disorder instances ${W_i}$ while sweeping $V$, the fidelity metric exhibits a collection of slightly displaced $\delta$ functions for the different realizations that display a first-order transition but a broad peak for a second-order one. For that reason, we use its typical value, shown in Fig.~\ref{fig:fid_susc} for $W=2$. A one-peak structure that systematically drifts to larger interactions as disorder is increased denotes that the typical fidelity susceptibility closely captures the onset of the Mott-CDW insulating phase when compared with the phase diagram in Fig.~\ref{fig:fig_1}(c1).

\subsection{Single-particle density of states}

\begin{figure}[t]
\centering

\includegraphics[width=0.99\columnwidth]{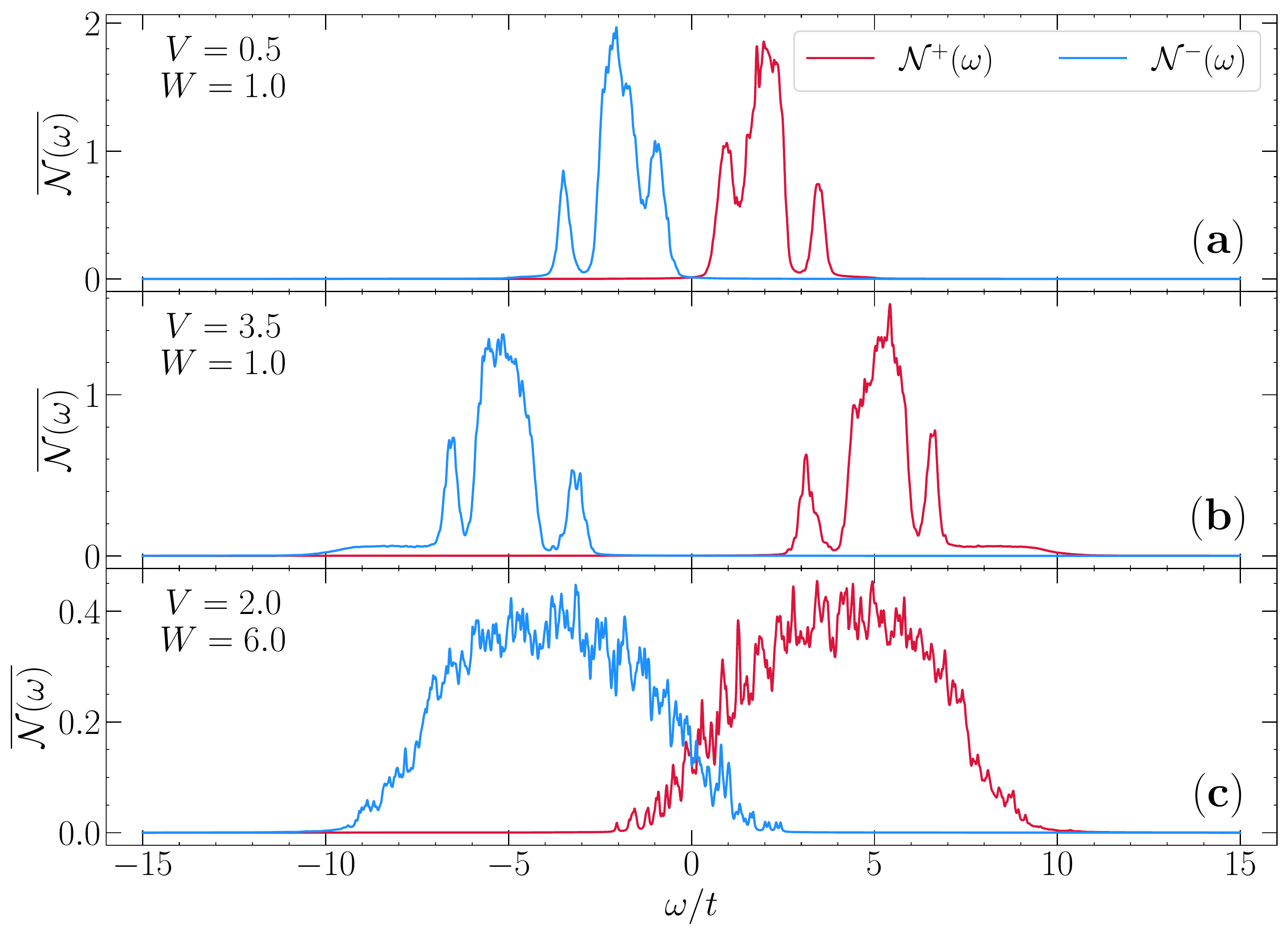}
\caption{The disorder-averaged single-particle density of states at three representative points of the phase diagram: (a) the Chern insulating, (b) Mott insulating, and (c) trivial Anderson insulating phases. All results are obtained as an average of 200 disorder realizations, using the staggered potential $\Delta = 0$ on the $18A$ lattice. A small Lorentzian broadening $\epsilon = 0.03$ is used to smooth the curves. The Anderson insulating phase is differentiated for being the only phase with a gapless structure in the DOS. In this case, insulating behavior ensues from an exponential localization of wave functions.
}
\label{fig:DOS_Delta0}
\end{figure}

A contrast between the different insulating phases can be drawn by comparing the single-particle DOSs. Unlike the two other insulating phases, the Anderson insulator is unique because it generally displays a gapless DOS, owing its insulating character to the localization of the wave functions. Figure~\ref{fig:DOS_Delta0} depicts the DOS at three points of the phase diagram for $\Delta = 0$ that correspond to each of the phases. Although it would be very instructive, a systematic study of the onset of the Anderson insulating phase via the gap closing in ${\overline{\cal N}(\omega)}$, which is consequently accompanied by a topological-to-trivial transition, is elusive since this single-particle gap is known to display finite-size effects which are much more dramatic than for the Chern number~\cite{Varney2011}.

\section{Summary}
The Haldane-Hubbard model is well characterized when disorder~\cite{Goncalves2018} and interactions~\cite{Varney2010,Varney2011} are separately included. By combining these two ingredients, we unveiled regimes of interacting topological Anderson insulators using numerically unbiased methods, showing their resilience to the presence of moderately large interactions before they render a trivial CDW insulating state. In principle, the survival of such a phase can be tuned to even larger interactions via the systematic enhancement of the next-nearest-neighbor hopping amplitude $t_2$, which is known for delaying the onset of the CDW Mott insulator~\cite{Varney2011}. Moreover, the characterization of the transition to the Mott-CDW phase, in which a first-order phase transition is replaced by a continuous one for sufficiently large disorder amplitude, was performed. Nonetheless, typical first-order phase transitions for certain realizations can still occur in finite systems with periodic boundary conditions, especially at small disorder amplitudes. While these results were obtained for relatively small system sizes, they point to a qualitative convergence when the lattice studied is increased. 

From an experimental point of view, the recent emulation of the Haldane model using ultracold atoms trapped in optical lattices~\cite{Jotzu2014} and the known flexibility to tune interactions that such settings provide~\cite{Chin2010} allow us to predict that similar physics can be experimentally verified with quantum emulators. In particular, effective nonlocal interactions similar to the type investigated here were also  recently demonstrated~\cite{Guardado2021}.

\begin{acknowledgments}
We acknowledge support from National Natural Science Foundation of China (NSFC) Grant No. NSAF-U1930402; R.M. acknowledges NSFC Grants No.~11974039 and No.~12050410263, and No.~12111530010. E.V.C.~ acknowledges partial support from FCT-Portugal through Grant No.~UIDB/04650/2020. Computations were performed on the Tianhe-2JK at the Beijing Computational Science Research Center.
\end{acknowledgments}

\appendix
\section{Chern number computation}\label{appendixA}
The calculation of the Chern number is based on the method described in Refs.~\onlinecite{Fukui05,Varney2010}, i.e., using a discretized form of Eq.~\eqref{eq:Chern}, by making use of twisted boundary conditions (TBCs)~\cite{Poilblanc1991, Niu1985}. When computing the many-body ground state in the torus $\{\phi_x,\phi_y\}$ of TBCs, the Berry curvature is written in terms of the normalized overlaps
\begin{align}
 U^{i,j}_x = \frac{\langle \psi_0^{i,j}|\psi_0^{i+1,j}\rangle}{|\langle \psi_0^{i,j}|\psi_0^{i+1,j}\rangle|}, && U^{i,j}_y = \frac{\langle \psi_0^{i,j}|\psi_0^{i,j+1}\rangle}{|\langle \psi_0^{i,j}|\psi_0^{i,j+1}\rangle|},
\end{align}
at consecutive points of the $\{i,j\}$ grid defined by the $\{\phi_x,\phi_y\}$ phases, when it is subdivided in $N_\alpha$ intervals, such as $\phi_\alpha = 2\pi i/N_\alpha$. Thus, the discretized Berry curvature assumes the form
\begin{equation}
 \tilde F_{i,j} = -{ i}\log\left(\frac{U_x^{i,j} U_y^{i+1,j}}{U_x^{i,j+1}U_y^{i,j}}\right),
\end{equation}
where the Chern number is written as $C = (1/2\pi)\sum_{i,j} \tilde F_{i,j}$.
We typically use $N_\alpha = 6$, which is sufficient to capture the correct Chern number for the different phases when $W=0$.

\section{Individual disorder realizations: Gap closing and phase diagrams}\label{appendixB}
\begin{figure}[t]
\centering
\includegraphics[width=0.99\columnwidth]{{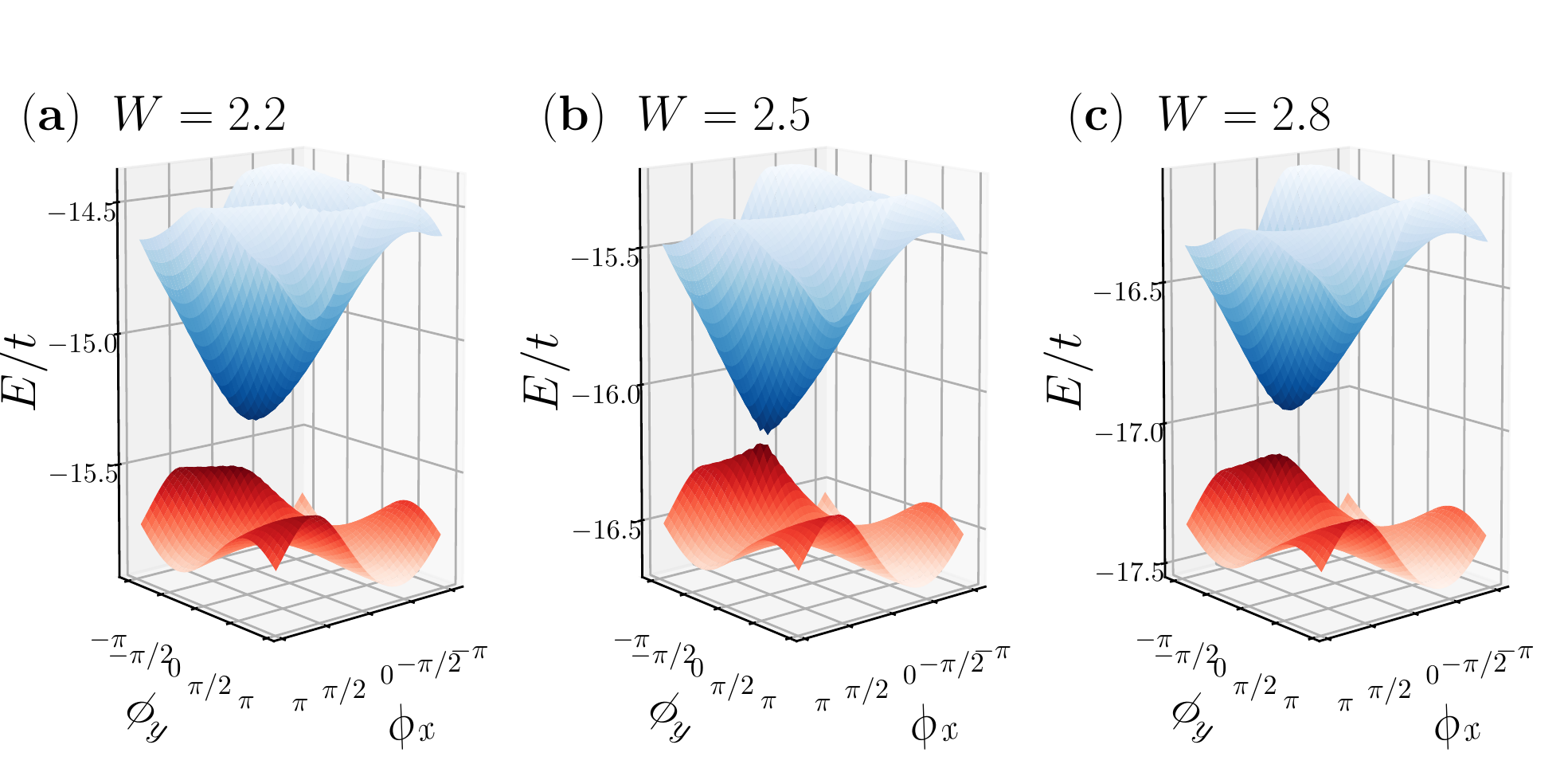}}
\caption{
Evolution of the eigenenergy surfaces ($E_0$ and $E_1$) in the torus of TBCs $\{\phi_x ,\phi_y\}$ for a fixed disorder realization and growing disorder amplitude (a) $W = 2.2$, (b) 2.5, and (c) 2.8. The interaction strength is $V = 0.5$, the staggered potential $\Delta=0$, and the lattice is $18A$. In (a), the computed Chern number is $C=1$, whereas in (c), $C=0$.
}
\label{fig:gap_closing_one_realz}
\end{figure}

An important remark is that for finite disorder, the statement that the gap closing condition occurs at one of the high-symmetry points $(\phi_x,\phi_y) = (0,0)\ (\pi, 0),\ (0, \pi)\ (\pi,\pi)$~\cite{Varney2011} is no longer valid given that point-group symmetries (in a special inversion) are now absent. As a result, the many-body gap closing, which underlies the change in topological properties of $|\psi_0\rangle$, can occur at any value of $(\phi_x ,\phi_y)$ on the patched torus. As an example, we show in Fig.~\ref{fig:gap_closing_one_realz} the eigenenergy surfaces for a single disorder realization $\{W_i\}$ when stretching the disorder amplitude. As for the clean case, a gap closing results in a change in the topological invariant, and its location in the TBC torus is now tied to the specific choice of the disorder configuration $\{W_i\}$. 

\begin{figure}[t]
\centering
\includegraphics[width=0.99\columnwidth]{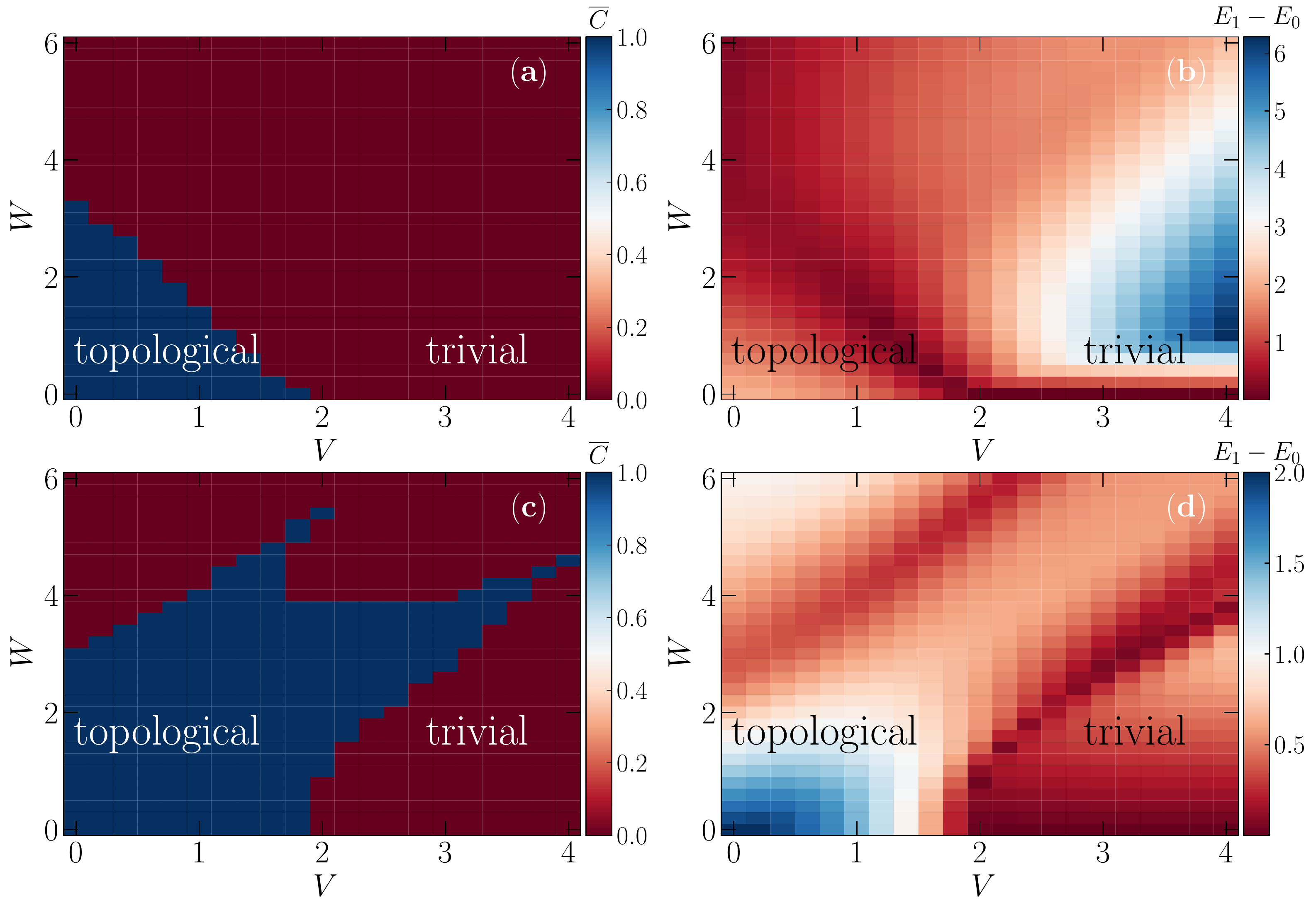}
\caption{
(a) and (c) One disorder realization instance of the Chern number phase diagram; 100 averages of such diagrams compose the ones in Fig.~\ref{fig:fig_1}. (b) and (d) The corresponding gap in the spectrum between the ground state and the first excited state for a fixed disorder landscape, calculated with periodic boundary conditions $(\phi_x,\phi_y) = (0,0)$. Here $\Delta=0.0$, and the lattice is $24A$.
}
\label{fig:one_realz_phase_diag}
\end{figure}

In particular, when constructing the phase diagrams, we noticed that a similar procedure results in easier convergence; that is, we focus on a single disorder realization to build one instance of the $W-V$ phase diagram, subsequently averaging the results for different realizations. In Figs.~\ref{fig:one_realz_phase_diag}(a) and ~\ref{fig:one_realz_phase_diag}(c), we show two such instances where the corresponding regions of topological nontrivial behavior are remarkably different, especially at large disorder amplitudes $W$. Nonetheless, we can directly interpret these results by observing the behavior of the many-body gap $E_1 - E_0$ in Figs.~\ref{fig:one_realz_phase_diag}(b) and ~\ref{fig:one_realz_phase_diag}(d). Apart from the $W=0$ line, where the system possesses quasidegenerate doublet states (related to odd and even CDW states with respect to parity), the change in the Chern number in the other regions of the phase diagram is traced again to a gap-closing condition. 

Last, we reemphasize that this gap closing may not occur for periodic boundary conditions $\phi_x=\phi_y=0$ in a finite lattice [such as the ones displayed in Figs.~\ref{fig:one_realz_phase_diag}(b) and \ref{fig:one_realz_phase_diag}(d)]. This can be observed for one of the disorder realizations (lower panels in Fig.~\ref{fig:one_realz_phase_diag}) as seen for interactions $2\lesssim V \lesssim 3$ and $W\simeq4$, where the Chern number changes but the many-body gap $E_1 - E_0$ is markedly finite. 

Yet in disordered and/or interacting systems, the discretized version of Eq.~\eqref{eq:Chern} can be used with just one single twist term corresponding to periodic boundary conditions, as long as large enough supercells are used~\cite{Zhang2013}. This is equivalent to the single ${\bf k}$-point formula for the Berry phase in calculations of the electrical polarization~\cite{Resta2010}. Therefore, the lack of gap closing for $\phi_x=\phi_y=0$ in the present case may also be interpreted as a finite-size effect due to the small system sizes used.

\begin{figure}[t]
\centering
\includegraphics[width=0.99\columnwidth]{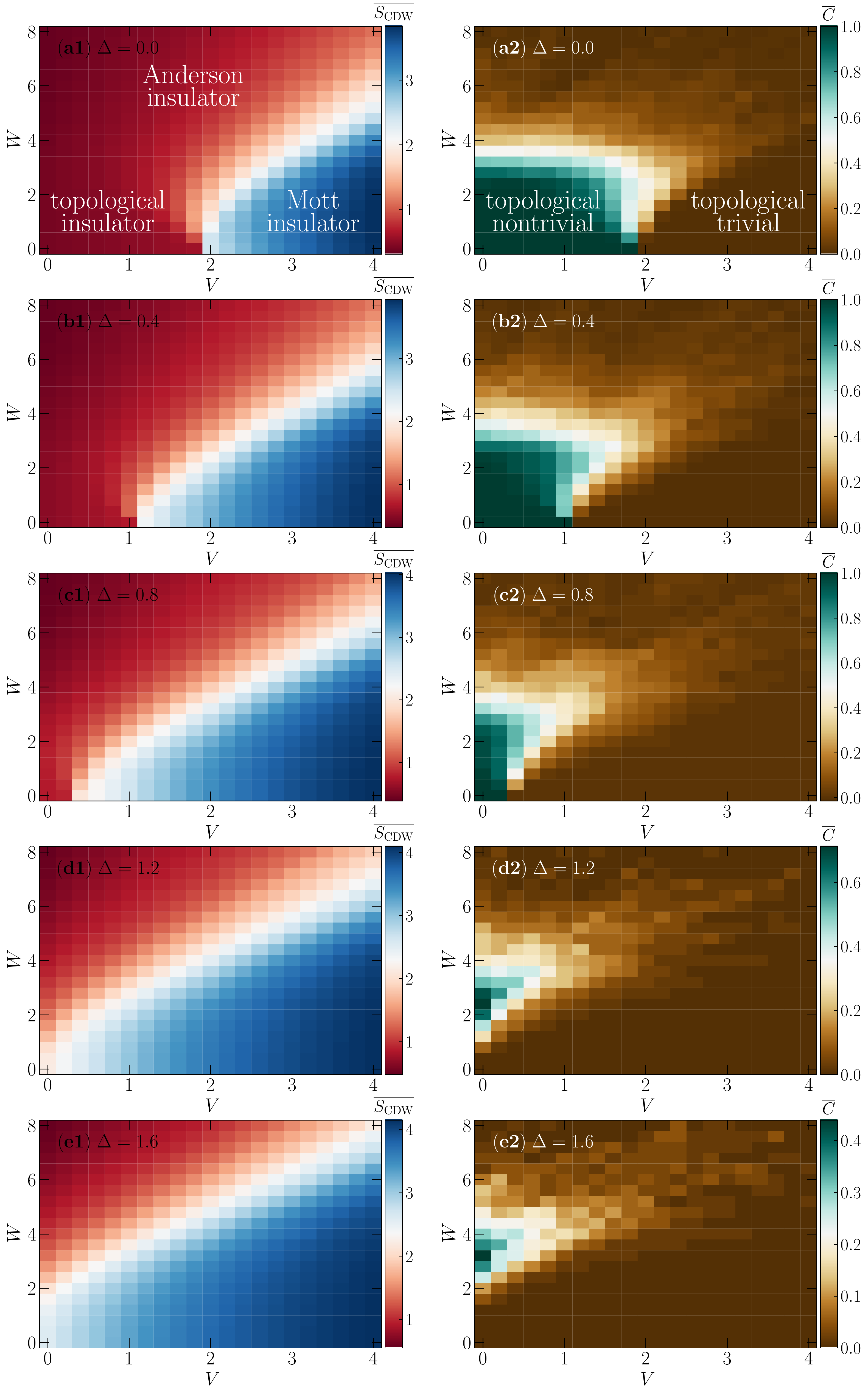}
\caption{
Evolution of the phase diagrams of the average CDW structure factor (left panels) and Chern number (right panels) with growing staggered potential $\Delta$ (from top to bottom, $\Delta = 0, 0.4, 0.8., 1.2$, and 1.6). Here the lattice used is $18A$.
}
\label{fig:18A_phase_diag}
\end{figure}

\section{Topological Anderson insulator with growing $\Delta$}\label{appendixC}
In the main text, we presented results for cases with both $\Delta = 0$ and $\Delta = 1.2$, arguing that in the latter, the topological Anderson insulating phase survives the inclusion of interactions. We now present a smooth interpolation between these two regimes in Fig.~\ref{fig:18A_phase_diag} for the $18A$ cluster. The CDW insulating phase displays a comparatively larger region in the phase diagram with a growing staggered potential, as one would expect, since both interacting and staggered potentials contribute to the formation of a charge ordered insulator. 

Even more interesting, one can see that the topological region, characterized by finite values of the average Chern number, can be seen to smoothly evolve until $\Delta_c = 3\sqrt{3}t_2 \simeq1.04$, beyond which the TAI, characterized as the regime in which nontrivial topology is recovered only in the presence of disorder, appears. The renormalization of the staggered trivial mass by $W$, as discussed in the Introduction, is seen to be possible even far from $\Delta_c$ [see Fig.~\ref{fig:18A_phase_diag}(e2)] and shows that an interacting topological Anderson insulator is still attainable in these regimes.

\bibliography{references}
\end{document}